\documentclass[12pt]{iopart}
\usepackage{graphicx}  
\begin{document}

\title[]{Space-time analysis of reaction at RHIC}

\author{Fabrice Reti\`ere\dag\   
\footnote[3]{To
whom correspondence should be addressed (fgretiere@lbl.gov)}
}

\address{\dag\ Lawrence Berkeley National Laboratory,
One Cyclotron Road, Berkeley CA 94720-8169}

\begin{abstract}

Space-time information about the Au-Au collisions produced at RHIC are
key tools to understand the evolution of the system and especially assess the
presence of collective behaviors. Using a parameterization of the system's final
state relying on collective expansion, we show that pion source radii can be tied
together with transverse mass spectra and elliptic flow within the same framework.
The consistency between these different measures provide a solid ground to
understand the characteristics of collective flow and especially the possible
peculiar behavior of particles such as $\Xi$, $\Omega$ or $\phi$. The validity
of the short time scales that are extracted from fits to the pion source size is also
addressed. The wealth of new data that will soon be available 
from Au-Au collisions at $\sqrt{s_{NN}} = $ 200 GeV, will provide a stringet test of the space-time analysis framework developped in these proceedings.

\end{abstract}



%\maketitle

\section{Introduction}

Ultra-relativistic heavy ions collisions are believed to produce 
initially hot systems that cool down by expanding until their temperature
is low enough to release hadrons. Were the conditions within these
systems adequate to free, for a time, quarks and gluons of their hadronic
confinement? This question will remain pending at the end of the proceedings.
Our aim will be to assess whether or not the particle space-time emission
pattern is consistent with a scenario where initially hot systems have cooled 
down by  expanding.  In this scenario, only a small fraction of the particles
measured in the detectors, are produced in the initial collisions between the  
nuclei. Most of the initially produced particles do not escape the system and reinteract inelastically with other particles. Only the high transverse
momentum particles are dominantly produced in the initial nucleus-nucleus collisions as
there is not enough energy available to produce them after the system
has cooled down. Following this observation, we only
consider particles with a transverse momentum smaller than 2 GeV/c.

Hydrodynamic models describe the evolution of systems from hot initial
stages until freeze-out by assuming zero mean free path, which provides the
limit of maximum transverse expansion~\cite{HydroGen}. However, this affirmation is not
strictly true because the equation of state, and especially its main
feature, the phase transition, regulates the pressure, i.e. the collective 
expansion strength. Hydrodynamic calculations may be used as a baseline
to assess the presence  of collective expansion, and ultimately
to extract the equation of state of the system. Complementary to 
the hydrodynamic calculations, we will also base our discussion on 
parameterization of the system final state. Such parameterizations allow
to quantify key points of the system final state as well as to investigate 
the consistency between various measures.

In these proceedings, we will first show that particle yield and
transverse momentum distribution
may be described assuming collective behaviors. Then, we will investigate
how the spatial extent of the pion source may be related to collective
expansion. Finally, we will discuss whether or not the time scales that 
are extracted assuming collective expansion make sense. 

\section{Collective flow and transverse momentum spectra}

The relative yields of many different particles species are well described
by parameterizations assuming that particles are statistically produced
within the available phase space~\cite{ChemFit}. With the additional assumption that the system is in thermal
equilibrium at freeze-out, the fit parameters may be understood as temperature
and chemical potential. The extracted temperature is on the order
of 170 MeV at RHIC energies, which is remarkably close to the phase 
transition temperature predicted by lattice QCD~\cite{LatQCDTc}. It suggests but 
does not prove that
particle yields are frozen out at the boundary between partonic and hadronic
matter.

Particle yields stop changing when the number of inelastic interactions 
become insignificant. However, the number of elastic or pseudo-inelastic 
interactions (such as
$\pi^+ \pi^- \rightarrow \rho \rightarrow \pi^+ \pi^-$) may still remain significant.  Thus, the 
particle momenta may still significantly change after the yields are frozen-out
(at chemical freeze-out) leading to a separate kinetic freeze-out stage.
Furthermore, if the chemical freeze-out corresponds to the boundary
between partonic and hadronic matter, the interactions  that lead to kinetic
freeze-out depend on hadronic cross-sections. Thus, the chemical and
kinetic freeze-out of particles with low hadronic cross-sections, such as,
presumably, $\Xi$, $\Omega$ or $\phi$ may coincide.

\begin{figure}
\begin{center}
\mbox{\includegraphics[width=0.6\textwidth]{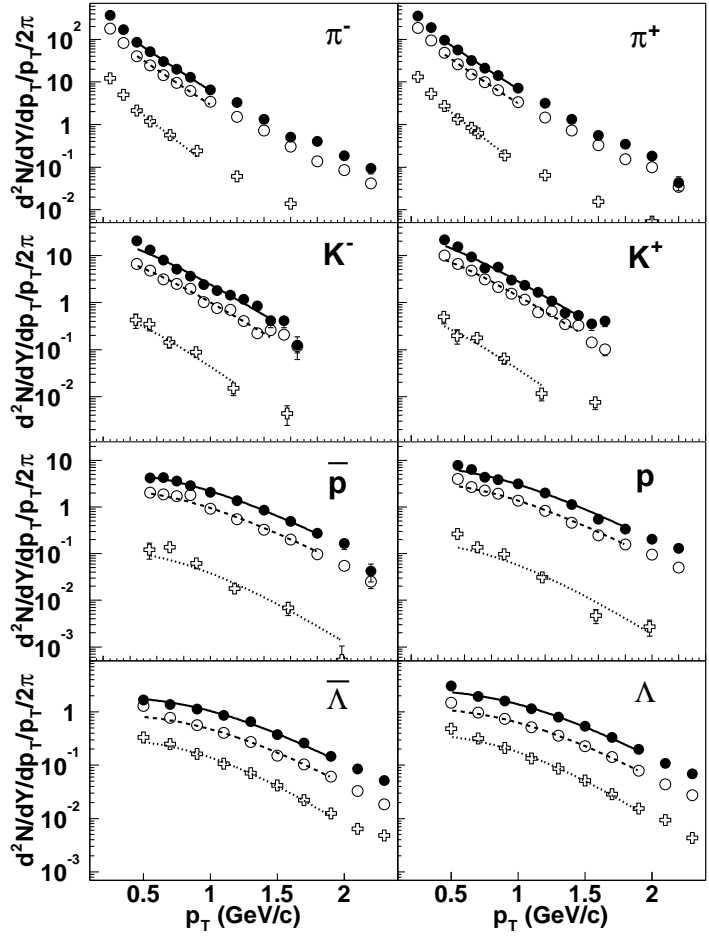}}
\caption{\label{fig:FitPtSpectra}Comparison of measured spectra in 
Au-Au collisions at $\sqrt{s_{NN}} = $ 130 GeV with the blast wave 
calculations performed with the best fit parameters in three centrality bins. 
Closed circle are central data,open circle are mid-central data and crosses are 
peripheral data.}
\end{center}
\end{figure}

Hydrodynamic calculations are successful at reproducing particle yields and transverse momentum spectra that have been measured at RHIC~\cite{HydroSpectra}. As shown in Figure~\ref{fig:FitPtSpectra}, the blast wave parameterization~\cite{SSH} is also successful at reproducing $\pi$, K, p, $\Lambda$ transverse momentum spectra measured in Au-Au collisions at $\sqrt{s_{NN}} = $ 130 GeV published by the PHENIX~\cite{PhenixSpectra130} and STAR~\cite{StarLaSpectra130} collaborations.  This parameterization is designed to mock up the final state of the hydrodynamic 
evolution. Thus, it relies on transverse expansion to reproduce
the spectra of different particle species with a single set of parameters:
a freeze-out temperature and a mean flow rapidity (or equivalently velocity). 
The freeze-out temperature is independent of the particle species the freeze-out time or position. The system is confined
within an infinitely long cylinder along the beam line. Longitudinal boost invariance
is assumed. 
The transverse flow rapidity
linearly increases from zero at the center to a maximum value 
($\rho_0 = 3/2 \langle \rho \rangle$) at the edge of the system.
In addition to the temperature and flow rapidity, the particle yields are 
also free parameters. The best fit parameters and $\chi^2/$dof are summarized
in table~\ref{tab:FitResult1}.

\begin{table}[t]
\centering \begin{tabular}{cccc}
\hline\hline
\multicolumn{1}{c}{} &
\multicolumn{1}{c}{Central}  &
\multicolumn{1}{c}{Mid-central} & 
\multicolumn{1}{c}{Peripheral} \\
\hline
$\pi$, K, p spectra~\cite{PhenixSpectra130} & 0-5\%  &15-30\% &60-92\%  \\
$\Lambda$ spectra~\cite{StarLaSpectra130}  & 0-5\% &20-35\% &35-75\% \\
\hline
$\chi^{2}$/dof          &  55.1/58 & 105.6/58 & 64.0/37 \\
T (MeV)                &  108 $\pm$ 3           & 106 $\pm$ 3          & 95 $\pm$ 4         \\
$ \langle \beta_{T} \rangle (c) $  &  0.53 $\pm$ 0.01     & 0.52 $\pm$ 0.02   & 0.47 $\pm$ 0.02 \\
\hline\hline
\end{tabular}
\caption{Upper section: data used in the fit with their different centrality range. Bottom section: bets fit parameters and $\chi^2/$ dof.}
\label{tab:FitResult1}
\end{table}

The kinetic freeze-ot temperature that is obtained from blast wave fits to transverse momentum spectra ($\approx$ 100 MeV) is significantly lower than the chemical freeze out temperature  ($\approx$170 MeV) . However, the study reported in Ref~\cite{Pols} shows that the 
separation between chemical and kinetic freeze-out vanishes if
resonance feed-down is properly accounted for. The authors of this paper
are able to reproduce the $\pi$, K and p transverse momentum spectra published by the PHENIX collaboration~\cite{PhenixSpectra130} with a temperature of 165 MeV
by decaying all the resonances that feed-down into  $\pi$, K and p (such as $\rho$, $\omega$, $K^*$, $\Delta$,...). However, this result is contradicted by a study
reported in Ref.~\cite{Peitzmann} where it is shown that forcing the chemical
and kinetic freeze-out temperature to coincide significantly increases the
$\chi^2$/dof. Thus, the effect of resonances on transverse momentum spectra
remains to be clarified in order to conclude whether freeze-out occurs in one
or two stages. In these proceedings, we will further investigate the later
hypothesis where chemical freeze-out precedes kinetic freeze-out.

One important consequence of a freeze-out in two stages is the possibility
that the $\Xi$, $\Omega$ and $\phi$ behave differently than $\pi$, K , p and 
$\Lambda$ because of their presumably lower hadronic cross-sections. In
that case their kinetic freeze-out temperature is expected to be close to the
chemical freeze-out temperature. Hence, if the chemical freeze-out stage 
coincides with the transition from partonic to hadronic matter, those particles
allow to measure the amount of collective expansion that build up at the
partonic stage. Such speculations can be investigated using the data
reported at this conference on $\Xi$, $\Omega$ and $\phi$ spectra
in Au-Au collisions at $\sqrt{s_{NN}} = $ 200 GeV. From 
$\Xi$, $\Omega$ and $\phi$ spectra measured in Au-Au collisions at 
$\sqrt{s_{NN}} = $ 130 GeV, blast wave fits allow to conclude only that 
$\Xi$ reach
kinetic freeze-out at a higher temperature and lower transverse flow velocity
than $\pi$, K, p and $\Lambda$~\cite{JaviersProc}. Due to the lack of statistics 
such claim cannot
be made from $\Omega$ or $\phi$ spectra. Interestingly, data reported by the
NA49~\cite{NA49MSB} and NA57~\cite{NA57MSB} collaborations show that
$\Xi$ and $\Omega$ spectra are well described by a blast wave calculation 
using the same parameters that reproduce $\pi$, K, p and $\Lambda$ spectra.
However, the statistical significance of these results remain to be addressed.

We have shown that particle yields and transverse momentum spectra fit in 
a scenario where the initial hot systems expand and cool down. On the other
hand, a few important issues have not been unambiguously established: does
the chemical freeze-out coincide with the partonic to hadronic matter 
transition? Do the hadronic cross-sections influence the temperature and
flow velocity at which particle freeze-out? Are the chemical and kinetic freeze-out stages decoupled? We will revisit this last question when discussing the time
scale issues. Before that, we will investigate the consequences of the collective
expansion on the position where particles are emitted. This study will allow
us to discuss a fundamental questions that we have ignored so far; is collective
expansion the only scenario that is consistent with RHIC data?

\section{Spatial issues, assessing collective flow}

Transverse momentum spectra do not show unambigously that the 
systems produced in ultra-relativistic heavy ion collisions undergo collective transverse expansions. Indeed, transverse mass scaling if it is established
may arise from models such as the Color Glass Condensate~\cite{CGC}. Furthermore, the NA49 collaboration has shown~\cite{InitRescat} that the transverse momentum distribution of protons produced by the projectile in p-Pb collusions is similar
to the distribution of protons from Pb-Pb collisions. Thus, initial state effects
such as Color Glass Condensate or random walk of partons may mock up
the effect of transverse flow. However, the collective expansion also 
affects the position of particle emission. Hence, measures which are
sensitive to the position may be used to assess the presence of collective
flow.

\begin{figure}[h]
\begin{center}
\mbox{\includegraphics[width=0.6\textwidth]{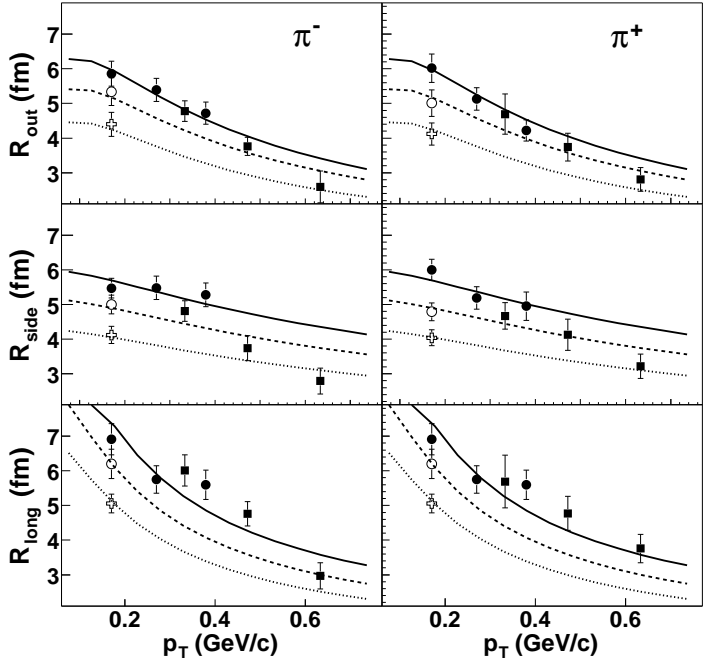}}
\caption{Comparison of the pion source data measured by the STAR (circles and crosses)~\cite{STARHbt} and the PHENIX (boxes)~\cite{PhenixHbt} collaborations in 
Au-Au collisions at $\sqrt{s_{NN}} = $ 130 GeV with the blast wave calculations.Only the STAR data were used to constrain the blast wave parameters. The closed circles are the central data,the open circles are the mid-central data and the crosses are peripheral data.
\label{fig:FitPionHbt}}
\end{center}
\end{figure}

Pion source sizes are measured by two-pion interferometry techniques. One
very important feature of this technique is that it probes the separation between
pions at low relative momentum. Thus when space-momentum correlations are
present, the measured source size does not reflect the whole source size. 
Space-momentum correlations are at the core of any collective expansion. Indeed,
in such scenario, particles push each other away from the hot center 
toward the vacuum. In other words, particles are emitted outward, 
their spatial and momentum azimuthal angles are close to each other. Furthermore, the particles that pick up a large momentum kick from the collective expansion
tend to be emitted close to the edge of the system. These features, which are
characteristic of the collective expansion, are obtained in hydrodynamic
calculations and parameterized in the blast wave framework.

Pion source sizes have been measured in Au-Au collisions at $\sqrt{s_{NN}} = $ 130 GeV by the STAR and PHENIX collaboration~\cite{STARHbt, PhenixHbt}.  The 
measured source radii shown in figure~\ref{fig:FitPionHbt}, are decomposed in the three directions, $r_{out}$, $r_{side}$ and $r_{long}$. They are calculated in the Longitudinally Comoving System where the pair velocity is zero. $r_{long}$ is parallel to the beam axis, $r_{out}$ is parallel to the
pair transverse momentum and $r_{side}$ is perpendicular to both beam axis
and pair transverse momentum. It can be shown that $r_{side}$ only probes
the spatial extent of the source while $r_{out}$ and $r_{long}$ are sensitive
to the system lifetime and the duration of particle emission~\cite{PrattHbt}.

\begin{table}[t]
\centering \begin{tabular}{cccc}
\hline\hline
\multicolumn{1}{c}{} &
\multicolumn{1}{c}{Central}  &
\multicolumn{1}{c}{Mid-central} & 
\multicolumn{1}{c}{Peripheral} \\
\hline\hline
$\chi^2/$dof & 10.9/15 & 0.7/3 & 0.6/3 \\
T (MeV) (fixed)               &  108            & 106          & 95        \\
$ \langle \beta_{T} \rangle $  (c) (fixed) &  0.53      & 0.52    & 0.47  \\
$ R (fm) $      & 12.9 $\pm$ 0.2        & 11.0 $\pm$ 0.4     & 9.1 $\pm$ 0.3 \\
$ \tau (fm/c) $      & 8.9 $\pm$ 0.3          &  7.4 $\pm$ 1.2      & 6.5 $\pm$ 0.8     \\ 
$ \Delta t (fm/c) $ & 0.002 $\pm$ 1.4        &   0.8 $\pm$ 3.2      &  0.8 $\pm$ 1.9 \\
\hline\hline
\end{tabular}
\caption{Blast wave parameters that best fit the pion source radii published 
by the STAR collaboration. The temperature and flow velocity parameters were
obtained by fit to transverse mass spectra (see table~\ref{tab:FitResult1}).}
\label{tab:FitResult2}
\end{table}

Hydrodynamic calculations fail to reproduce the measured radii~\cite{HydroHbt}. In most cases, these calculations underestimate $r_{side}$ and overestimate 
$r_{out}$ and $r_{long}$. In other words, they underestimate the system size
and over estimate its lifetime and emission duration. Does it mean that the
collective expansion scenario is ruled out? Not necessarily, since the transverse
momentum dependence of the radii is, in general, well described and it is only the magnitude of the radii that cannot be reproduced. This observation is confirmed
when using the blast wave parameterization as shown in figure~\ref{fig:FitPionHbt}. In this case,
the radii are well reproduced because in the blast wave parameterization
the system size, life time and emission duration are tunable parameters. The
values of the best fit parameters are shown in table~\ref{tab:FitResult2}. The exact same parameterization that was used to the fit the transverse momentum spectra is applied~\cite{BlastWaveHbt}. The 
temperature and the flow velocity have been fixed to the values obtained from
fits to transverse momentum spectra. The remaining free parameters are the 
radius (R) of the cylinder that confines the system, the system proper time 
($\tau = \sqrt{t^2-z^2}$) and the emission duration ($\Delta t$). The longitudinal boost
 invariance assumption motivates the use of the parameter $\tau$ rather than
the time, t, in the laboratory frame.

The good fit of the data obtained with the blast wave parameterization shows
that transverse momentum spectra and pion source size can be interpreted 
consistently in terms of collective expansion.  Furthermore, preliminary data from the STAR 
collaboration shows that the blast wave parameterization predicts 
an average space-time separation between pion, kaon and proton sources that 
is consistent with the data~\cite{BlastWaveNonId}. The list of measures that
can be used to test the blast wave parameterization also includes pion source radii measured with respect to the reaction plane~\cite{STARHbtAsRP}, and kaon source radii. Thus, collective transverse expansion as parameterized in the blast
wave framework provide a consistent, well constrained description
of the final state of the systems produced in Au-Au collisions at RHIC. Bringing in
two-particle correlation analyses has allowed us to assess  the presence of collective
expansion.

However, this definite conclusion may not hold when new and more precise
data are available from Au-Au collisions at 200 GeV. For example, the invariant
radius extracted from two $K^0_s$ interferometry reported by the STAR collaboration~\cite{SelemonK0s} is too large to be reproduced in the blast wave framework. On the other hand, the blast wave parameterization may prove
to be a very valuable tool to assess whether $\Xi$, $\Omega$ and $\phi$
undergo the same collective expansion than $\pi$, K, p and $\Lambda$, since it
can be used to interpret  transverse momentum spectra as well as $v_2$ and
results from two-particle correlation analysis such as $\pi^{\pm}-\Xi^{-}$
 or $\pi^{\pm}-\Omega^{-}$.

\section{A problem with the time scales?}

The agreement of the blast wave parameterization with data is reached because
the system lifetime and emission duration are small. Hydrodynamic calculations
are unable to produce such a short life-time. But, do such short time scales make
sense? Only extreme models achieve short time scales 
(~\cite{Humanic, MolnarHbt} for example). Thus, we will investigate whether
or not these short time-scales are supported by other measurements.
 
\begin{figure}[h]
\begin{center}
\mbox{\includegraphics[width=0.8\textwidth]{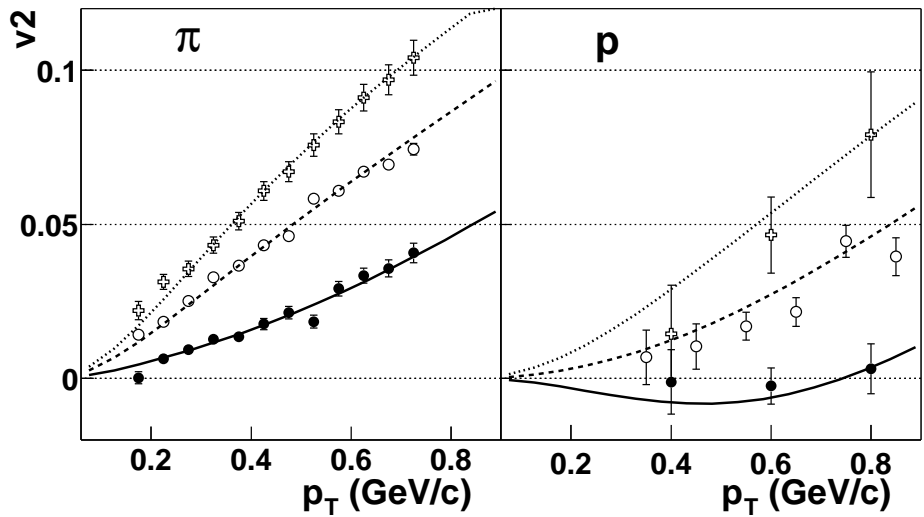}}
\caption{Comparison of $v_2$ for pions and protons measured by the 
STAR collaboration~\cite{STARV2} in 
Au-Au collisions at $\sqrt{s_{NN}} = $ 130 GeV
with the blast wave calculation obtained
with the parameters that best fit the data. The crosses are the most peripheral
events (45-85\% of centrality), the open circle at the mid-peripheral events 
(11-45\%) and the closed circles are the central data (0-11\%).
\label{fig:FitV2}}
\end{center}
\end{figure}

The spatial source shape carries qualitative information about the system lifetime. 
When the colliding ions do not fully overlap, the initial energy density is distributed
over an ellipse. The spatial energy density gradients are larger along the ellipse short radius (in-plane radius) than along the long radius (out-of-plane radius), which gives rise to pressure gradient stronger in-plane than out-of-plane. This phenomenon is experimentally quantified by the parameter $v_2$ as an 
azimuthal anisotropy of the particle emission, where more particles are emitted 
in-plane than out-of-plane~\cite{STARV2}. Furthermore, the collective expansion
fights against the initial anysotropy of the system. Along its lifetime the system
evolves from an out-of-plane extended shape towards a cylindrical shape and even
towards an in-plane extended shape depending on how long it lives~\cite{HydroShape}. Thus, a short life time
would be associated with an out-of-plane extended source while a long life time
would be associated with an in-plane source.  The blast wave parameterization may be used to 
extract the source shape from the data by adding two additional parameters 
allowing the final state of the system to be an ellipse instead of a disk and by
allowing the flow velocity to vary with the azimuthal angle. The ellipse shape
is parameterized by splitting the system radius in two: a radius in-plane and
a radius out-of-plane. Figure~\ref{fig:FitV2} shows a fit to the $v_2$ parameters performed
with the blast wave parameterization by fixing the temperature and overall 
flow velocity ($\langle \beta_T \rangle$) to the values obtained by fits to transverse
momentum spectra. From these fits, we extract the source aspect ratio (in-plane radius divided by the out-of-plane radius). Table~\ref{tab:FitResult3}  shows a comparison of
the initial aspect ratio obtained by a simple Wood-Saxon nuclear overlap calculation~\cite{Glauber} with 
the final aspect ratio obtained by blast wave fits to $v_2$. The important point
to notice is that the source remains out-of-plane extended, which is consistent
with short lived systems. The same qualitative conclusion has been independently obtained
by analysis of the preliminary STAR data of the pion source size with respect
to the reaction plane angle~\cite{STARHbtAsRP}. The short system life time
obtained from analysis of the pion radii and the out-of-plane extended source of the system final state are consistent.

\begin{table}[t]
\centering \begin{tabular}{cccc}
\hline\hline
\multicolumn{1}{c}{} &
\multicolumn{1}{c}{Central}  &
\multicolumn{1}{c}{Mid-central} & 
\multicolumn{1}{c}{Peripheral} \\
\hline\hline
$\chi^2/$dof & 14.6/13 &47.4/16 & 9.7/13 \\
\hline
T (MeV) (fixed)               &  108            & 106          & 95        \\
$ \langle \beta_{T} \rangle $  (c) (fixed) &  0.53      & 0.52    & 0.47  \\
$ \langle \beta_{T in-plane} \rangle / \langle \beta_{T out-of-plane} \rangle   $      & 1.067 $\pm$ 0.009          &  1.060 $\pm$ 0.007      & 1.05 $\pm$ 0.01     \\ 
$ R_{in-plane} / R_{out-of-plane} $      & 1.01 $\pm$ 0.03        & 0.86 $\pm$ 0.06     & 0.79 $\pm$ 0.05 \\
\hline
 \begin{tabular}{c}
Inital state\\
$ R_{in-plane} / R_{out-of-plane} $ 
\end{tabular}
    & 0.80 & 0.59 & 0.42\\
\hline\hline
\end{tabular}
\caption{Blast wave parameters that best fit the pion and proton $v_2$. 
The last row is a calculation of the source aspect ratio from 
a nuclear overlap approximation~\cite{Glauber}.
}
\label{tab:FitResult3}
\end{table}

Is the very short emission duration extracted from pion source radii also 
consistent with other measures? Before discussing this issue, it is important
to recall that the emission duration is mostly determined by the difference
(or the ratio) between the $r_{side}$ and $r_{out}$. As pointed out by the 
CERES collaboration~\cite{CeresHbt}, the ratio $r_{out}/r_{side}$ and hence the emission duration may be artificially increased by a inappropriate Coulomb correction applied to the two-pion correlation functions. With this point in mind, we argue that 
emission duration cannot be very small (i.e. less than 1 fm/c) because it wouldn't
leave enough time for the system to cool down from chemical to kinetic 
freeze-out and because it is not consistent with the measured yield of 
$K^*$~\cite{STARKStar} and 
$\Lambda^*(1520)$~\cite{STARLambdaStar} resonances. The first point is valid if
the chemical freeze-out temperature is truly higher than the kinetic freeze-out
temperature. In this case, the requirement that the entropy cannot decrease 
imposes, in a bounded system such as the one described by the blast wave parameterization, that the time between chemical and kinetic freeze-out is larger than
4-5 fm/c~\cite{OlgaFuqiang}.  The second point has to do with the fact that
the measured $K^*$ and $\Lambda^*(1520)$ yields are lower than calculated at
chemical freeze-out using the same parameters (temperature and chemical potential) that reproduce the relative yield of all the other hadrons. This 
suppression is understood by  arguing that the $K^*$ and $\Lambda^*$ decay products lose the invariant mass correlation when reinteracting (pseudo-)elasticaly before freezing-out. These data challenge the scenario
where chemical and kinetic freeze-out coincide, as this
interpretation is only valid if (pseudo)elastic interactions occur after chemical freeze-out.
The lifetime of the $K^*$ and
$\Lambda^*$ provide a gauge of the time between chemical and kinetic 
freeze-out which is on the order of 4-5 fm/c~\cite{STARLambdaStar}. Thus, the short emission
duration extracted from fit to pion radii is inconsistent with other measures. 
However, this conclusion will need to be revisited when pion source radii
are extracted using the proper Coulomb correction technique.

\section{Conclusions}

Assessing the presence and characteristics of collective behaviors in 
ultra-relativistic heavy ion collisions is a key step towards the discovery of
partonic matter. This step is well underway; we have shown that the data 
from Au-Au collisions at $\sqrt{s_{NN}} = $ 130 GeV are consistent with systems undergoing a collective expansion. However, definite conclusions await the  resolution of the following pending issues:

\begin{itemize}
\item Carefully assess the effect of resonance feed-down on transverse momentum spectra.
\item Improve the two-pion interferometry data by performing a proper Coulomb correction and extending the data range to higher transverse momentum. This may clarify the time scale issues.
\item Establish in a statisticaly significant manner whether or not $\Xi$, $\Omega$ 
and $\phi$ freeze-out at the same temperature and transverse flow velocity 
as $\pi$, K, and protons.
\item Investigate the no flow hypothesis by studying p-p and d-Au collisions. Indeed, space-momentum correlations such as the one that arrise from jet fragmentation may mock up the effect of flow.
\end{itemize}

In addition to the measure described in these proceedings, new analyses will 
available soon. The large statistics available at RHIC will
allow to study the behavior of $\Xi$, $\Omega$  and $\phi$, with never used
before tools such as $v_2$ or two-particle correlations. Balance function
analyses may also bring crucial information. Furthermore the
development of parameterizations such as the blast wave provides new
opportunities to assess the consistency of the data within a single framework. However, with the quality and the variety of the data increasing significantly such
parameterizations will be very significantly challenged in the coming future. In any
cases, combining space-time analysis with yield and spectra analysis will remain the key to reach a global understanding of the ultra-relativistic heavy ion collisions.
\\
I wish to thank the conference organizers for inviting me and hence providing me with the opportunity of developing these ideas. The blast wave fits were developed together with Mike Lisa following ten years of development within the community. These proceedings is the results of fruitfull discussions with 
Ulrich Heinz,  Mike Lisa, Dan Magestro, Thomas Peitzmann, Kai Schweda, 
Nu Xu,  and Zanghbu Xu, Eugene Yamamoto. I also wish to thank the whole STAR collaboration for the stimulating discussions that have gone along the release of the preliminary data shown at this conference.
\\

\end{document}